\documentclass{aastex7}
\usepackage{amsmath,amssymb}

\begin{document}

\title{Intermediate-mass-ratio inspirals with general dynamical friction in dark matter minispikes}
\correspondingauthor{Hong-Bo Jin}
\author{Yu-Chen Zhou}
\email{zhouyuchen21@mails.ucas.ac.cn}
\affiliation{The School of Physical Sciences, University of Chinese Academy of Sciences, Beijing 100049, China}
\affiliation{The International Centre for Theoretical Physics Asia-Pacific, University of Chinese Academy of Sciences, Beijing 100190, China}
\author{Hong-Bo Jin}
\email[show]{hbjin@bao.ac.cn}
\affiliation{The International Centre for Theoretical Physics Asia-Pacific, University of Chinese Academy of Sciences, Beijing 100190, China}
\affiliation{National Astronomical Observatories, Chinese Academy of Sciences, Beijing 100101, China}
\author{Cong-Feng Qiao}
\email{qiaocf@ucas.ac.cn}
\affiliation{The School of Physical Sciences, University of Chinese Academy of Sciences, Beijing 100049, China}
\affiliation{The International Centre for Theoretical Physics Asia-Pacific, University of Chinese Academy of Sciences, Beijing 100190, China}
\author{Yue-Liang Wu}
\email{ylwu@itp.ac.cn}
\affiliation{The International Centre for Theoretical Physics Asia-Pacific, University of Chinese Academy of Sciences, Beijing 100190, China}
\affiliation{Institute of Theoretical Physics, Chinese Academy of Sciences, Beijing 100190, China}

\date{\today}

\begin{abstract}

The intermediate-mass-ratio inspirals (IMRIs) may be surrounded by dark matter (DM) minispikes. The dynamical friction from these DM minispike structures can affect the dynamics and the gravitational wave (GW) emission of the IMRIs. We analyze the effects of general dynamical friction, with a particular contribution from DM particles moving faster than the stellar-mass black hole in an eccentric IMRI. 
Our calculation show that these DM particles tends to eccentricify the orbit, therefore the evolution of the eccentricity depends on the competition between the fast moving DM particles and the slow moving DM particles. The results show that the dynamical friction enhances the eccentricity when $\gamma_\mathrm{sp}\lesssim2.0$, and the general dynamical friction is able to increase the eccentricity. We also analyze the effects of general dynamical friction on the GW characteristic strain. 
The results indicate that the characteristic strain is suppressed at lower frequencies, and the peak value of the characteristic strain occurs at higher frequencies as the power law index of DM minispike $\gamma_\mathrm{sp}$ increases. For the first time, a relation between the frequency peak value of characteristic strain of GWs and $\gamma_\mathrm{sp}$ is established. Using this analytical relation, the presence of DM and its halo density may be determined potentially from future GW data.
\end{abstract}

\section{Introduction}
\label{s1}
The intermediate-mass black holes (IMBHs), with masses $10^2 M_{\odot}\lesssim M \lesssim 10^5 M_{\odot}$ between stellar-mass (below $\sim100 M_\odot$ ) and supermassive black holes (BHs) (above $\sim 10^5 M_\odot$), have been found by LIGO-VIRGO collaboration in the third observing run (O3)of the gravitational wave(GW). There are eight GW signals with a binary merger of mass above $\sim100 M_\odot$\cite{LIGOScientific:2021tfm, LIGOScientific:2021usb}. 
In the observed GW events, the merger maximum mass of two black holes reached about $185 M_{\odot}$, whose event name is GW190426\_190642\cite{LIGOScientific:2021usb}.

An IMBH may be surrounded by a stellar mass compact object, such as a black hole, and forms an intermediate-mass-ratio inspiral (IMRI) system, which may also be immersed in a halo of dark matter (DM). IMRIs also emit GWs and the stronger ones have the low frequency below 10 Hz.  The dynamics and GW emission of IMRIs is affected by the dynamical friction from the DM minispike \cite{1943ApJ....97..255C}, leading to a faster inspiral and a dephasing effect of GW signal \cite{Eda:2013gg, Eda:2014kra}. Such effects on GW signals may be detected by the future space-borne GW detectors such as LISA \cite{2022NatAs...6.1334B}, Taiji \cite{Hu:2017mde} and TianQin \cite{TianQin:2015yph}.The effects of dynamical friction in eccentric IMRIs are investigated in the paper\cite{PhysRevD.100.043013}, which is a more realistic description for IMRIs. Their results show that the dynamical friction tends to increase the eccentricity of the orbits. Other effects such as accretion \cite{PhysRevD.97.064003,Dai:2021olt,Becker:2022wlo}, halo feedback \cite{PhysRevD.102.083006} and relativistic effects \cite{Traykova:2021dua,Speeney:2022ryg} are also investigated in previous studies. The effects of dynamical friction on GW spectra and signal-to-noise ratios (SNRs) are investigated in \cite{Li:2021pxf}.

The inclusion of relative velocities between DM particles and the stellar-mass BH (called as the secondary BH $m_2$ in a IMRI) in the dynamical friction was considered in \cite{PhysRevD.105.063029}. By adopting the phase space description of DM halos, they find that the dynamical friction tends to decrease the eccentricity of the orbits. They also suggested that the circularization rate of IMRIs can be a new probe of the DM minispike. The contribution from the particles moving faster than the secondary BH was studied in \cite{dosopoulou2023dynamical}. They found that in this case the dynamical friction tends to decrease the eccentricity of the orbits under the condition of the spike power law index $\gamma_\text{sp} \gtrsim 1.8$. A significantly enhanced dephasing effects is also presented in their results.

In this paper, we focus on the impact of dynamical friction effects on the orbital evolution in IMRI systems. Our numerical results demonstrate that dynamical friction may lead to an increase in orbital eccentricity. Specifically, our calculation show that dynamical friction caused by more rapidly moving DM particles tends to increase the orbit's eccentricity. Consequently, dynamical friction contributes to this eccentricity enhancement.

Additionally, we explore the influence of these effects on the characteristic strain of GWs. We find that the characteristic strain is suppressed at lower frequencies, and the peak value of characteristic strain occurs at higher frequencies as the power law index of DM minispike increases. Finally, we determine a general relationship between the frequency peak value of characteristic strain of GWs and the power law index of DM minispike. This relationship can be utilized to constrain the power law index of the DM minispike. 
It implies that using the analytical relations, the the presence of dark matter, as well as the dark matter halo density distribution, can be determined from the detected GW data in the future.

This paper is organized as follows. In Sec. \ref{s2} we introduce the general dynamical friction and dynamical equations in IMRIs.  In Sec. \ref{s3A} we study the dynamical friction effects on orbital evolution of IMRIs. In Sec. \ref{s3B} we study the dynamical friction effects on GW characteristic strain. In Sec. \ref{s5} we give our conclusion. 
  
\section{Methods}
\label{s2}
\subsection{General dynamical friction and dynamical equations in IMRIs}
\label{s2A}
We consider an IMRI where the stellar mass object is a BH. The central IMBH is surrounded by a DM minispike, which has a static, spherically symmetric distribution with a simple power law \cite{PhysRevD.105.043009}
\begin{equation}
\label{eq1}
    \rho_{\textrm{DM}}\left(r\right)=
\left\{
\begin{array}{ll}
	\rho_\mathrm{sp}\left(\frac{r_\mathrm{sp}}{r}\right)^{\gamma_{\textrm{sp}}}, 
	&r_{\textrm{in}}\le r \le r_{\textrm{sp}}\\
	
	0, &r<r_{\textrm{in}}
\end{array}
\right.
,
\end{equation}
where r is the distance from the central IMBH, $r_{\textrm{in}}=4m_1$ is the inner radius \cite{PhysRevD.88.063522} and $r_{\textrm{sp}}=0.54$ $\mathrm{pc}$ is the maximal radius of the DM spike. $\rho_\mathrm{sp}=226M_\odot/\mathrm{pc}^3$ is the density at $r_\mathrm{sp}$. The values of power law index depends on the initial DM halo. For example, the initial DM halo with a NFW profile corresponds to $\gamma_{\textrm{sp}}=7/3$. In this paper, the range of $\gamma_{\textrm{sp}}$ is $0.5<\gamma_{\textrm{sp}}<3$.

The density of DM particles can also be expressed through the phase-space distribution function
\begin{equation}
\label{eq2}
\rho_{\textrm{DM}}(r)=4\pi\int^{v_\mathrm{esc}}_0{v^{\prime 2}f(\mathcal{E})dv^\prime}
\end{equation}
where $v_\mathrm{esc}$ is the escape velocity of the system. The phase-space distribution function $f(\mathcal{E})$ can be obtained by Eddington’s formula
\begin{equation}
\label{eq3}
f(\mathcal{E})=\frac{\rho_\mathrm{sp}}{(2\pi)^{3/2}}\left(\frac{r_\mathrm{sp}}{m_{1}}\right)^{\gamma_{\mathrm{sp}}}\frac{\Gamma(\gamma_{\mathrm{sp}}+1)}{\Gamma(\gamma_{\mathrm{sp}}-\frac{1}{2})}\mathcal{E}^{\gamma_{\mathrm{sp}}-3/2}
\end{equation}
where $\mathcal{E}=\Psi(r)-\frac{1}{2}v^2$, $\Psi(r)=\frac{m_1}{r}$ is the relative Newtonian gravitational potential close to the central IMBH. The range of power law index in Eq.(\ref{eq1}) is given by the gamma function and the requirement of finite enclosing mass in this expression.

Dynamical friction leads to the energy and angular momentum loss of the binary system. The general expression of dynamical friction is \cite{1943ApJ....97..255C,PhysRevD.105.063029}
\begin{equation}
\label{eq4}
F_{\mathrm{DF}}=\frac{4 \pi m_{2}^{2} \rho_{\mathrm{DM}}(r)\xi(v)}{v^{2}}
\end{equation}
Here we adopt the similar expression in \cite{PhysRevD.105.063029}, where $\xi(v)$ denotes the DM particles moving with different velocities relative to the secondary BH. However, the contribution of DM particles moving faster than $v$, the velocity of secondary BH, is also included in our formula. $\xi(v)$ can be determined through the following equation
\begin{equation}
\label{eq5}
\begin{aligned}
\rho_\mathrm{DM}(r)&\xi(v)=4\pi \ln{\Lambda}\int^v_0{v^{\prime 2} f\left(\mathcal{E}\right)dv^\prime}\\
&+4\pi\int^{v_\mathrm{esc}}_v{v^{\prime 2}f\left(\mathcal{E}\right)
\left[\ln{\left(\frac{v^\prime+v}{v^\prime-v}\right)}-2\frac{v}{v^\prime}\right]dv^\prime}
\end{aligned}
\end{equation}
where $\ln\Lambda$ is the Coulomb logarithm, and we adopt the value in \cite{PhysRevD.102.083006}, $ \ln \Lambda= \ln \sqrt{m_1/m_2}$.

GW emission also leads to the energy and angular momentum loss of the IMRIs, and reduces the eccentricity of the IMRI orbit \cite{Maggiore:2007ulw,PhysRev.131.435}. The total energy and angular momentum loss include both dynamical friction and GW emission
\begin{gather}
\frac{dE}{dt}=\left\langle\frac{dE}{dt}\right\rangle_{\mathrm{GW}}+\left\langle\frac{dE}{dt}\right\rangle_{\mathrm{DF}}\label{eq6}\\
\frac{dL}{dt}=\left\langle\frac{dL}{dt}\right\rangle_{\mathrm{GW}}+\left\langle\frac{dL}{dt}\right\rangle_{\mathrm{DF}}\label{eq7}
\end{gather}
where $\langle\rangle$ denotes the time average over one orbit period under the assumption of adiabatic approximation.

The evolution equations for semimajor axis $a$ and eccentricity are \cite{PhysRevD.105.043009}
\begin{gather}
\frac{da}{dt}=\frac{dE}{dt} / \frac{\partial{E}}{\partial{a}} =\frac{2a^2}{M\mu}\frac{dE}{dt}\label{eq8}\\
\frac{de}{dt}=-\frac{1-e^2}{2e}\left(\frac{dE}{dt}/E+2\frac{dL}{dt}/L\right)\label{eq9}
\end{gather}
where $M$, $\mu$ is the total mass and the reduced mass of IMRIs respectively. The evolution of semimajor axis $a$ and eccentricity $e$ can be obtained by solving Eq. (\ref{eq8}) and Eq. (\ref{eq9}) numerically. We use the publicly available \textsc{IMRIpy} code \cite{PhysRevD.105.043009,Becker:2022wlo} and modify several expressions of the code.
\subsection{gravitational characteristic strains and sensitivity curves}
GWs will be emitted during the inspiral stage. Here we adopt the gravitational waveforms to the leading post-Newtonian order \cite{Martel:1999tm} and the Fourier transform of GW signal is \cite{Moore_2018}
\begin{equation}
\begin{aligned}
\label{eq10}
\tilde{h}_{+, \times}^{(n)}(f)= & -\frac{\mathcal{M}}{2 D_{L}} \frac{\left(2 \pi \mathcal{M} F\left(t_{n}^{*}\right)\right)^{2 / 3}}{\sqrt{n \dot{F}\left(t_{n}^{*}\right)}} \\
& \times\left[C_{+, \times}^{(n)}\left(t_{n}^{*}\right)+i S_{+, \times}^{(n)}\left(t_{n}^{*}\right)\right] e^{i \psi_{n}},
\end{aligned}
\end{equation}
where $\mathcal{M}=\mu^{3/5}M^{2/5}$ is the chirp mass, $F$ is the mean orbital frrequency,  $C_{+, \times}^{(n)}$, $S_{+, \times}^{(n)}$ are the harmonic coefficients and $t_n^*$ is the stationary time of each harmonic. The frequency of n-th harmonic ranges from the initial frequency of the system $nF_0$ to the final frequency of the innermost stable circular orbit $nF_\text{isco}$.

The dimensionless characteristic strain is \cite{Moore:2014lga}
\begin{equation}
\label{eq11}
h_\mathrm{c}(f)=2f\lvert\tilde{h}(f)\rvert
\end{equation}

The sensitivity curves of space-borne GW detectors like LISA and TAIJI are \cite{Robson:2018ifk, Liu:2023qap}
\begin{equation}
\label{eq12}
S_{n}(f)=\frac{10}{3 L^{2}}\left(P_{\mathrm{OMS}}(f)+\frac{4 P_{\mathrm{acc}}(f)}{(2 \pi f)^{4}}\right)\left(1+\frac{6}{10}\left(\frac{f}{f_{*}}\right)^{2}\right)+S_{c}(f)
\end{equation}
where $P_{\mathrm{OMS}}(f)$, $P_{\mathrm{acc}}(f)$ and $S_{c}(f)$ are optical metrology noise, test mass acceleration noise and confusion noise respectively.

In order to compare with the characteristic strain, the sensitivity curve is rewritten in terms of characteristic strain.
\begin{equation}
\label{eq13}
h_\mathrm{n}(f)=\sqrt{fS_\mathrm{n}(f)}
\end{equation}
The characteristic strains and sensitivity curves can now be plotted in one figure and the detectability of characteristic strain can be discussed.

\section{Results}
\label{s3}
\subsection{Dynamical friction effects on orbital evolution}
\label{s3A}

First we look at the dynamical friction effects on orbital evolution. For the purpose of comparison, we set the mass of central IMBH to be $m_1=1000\ M_\odot$, the mass of secondary BH to be $m_2=10\ M_\odot$, which is the same to \cite{PhysRevD.100.043013,PhysRevD.105.063029}.

\begin{figure*}[htbp]
        \includegraphics[width=0.3\textwidth]{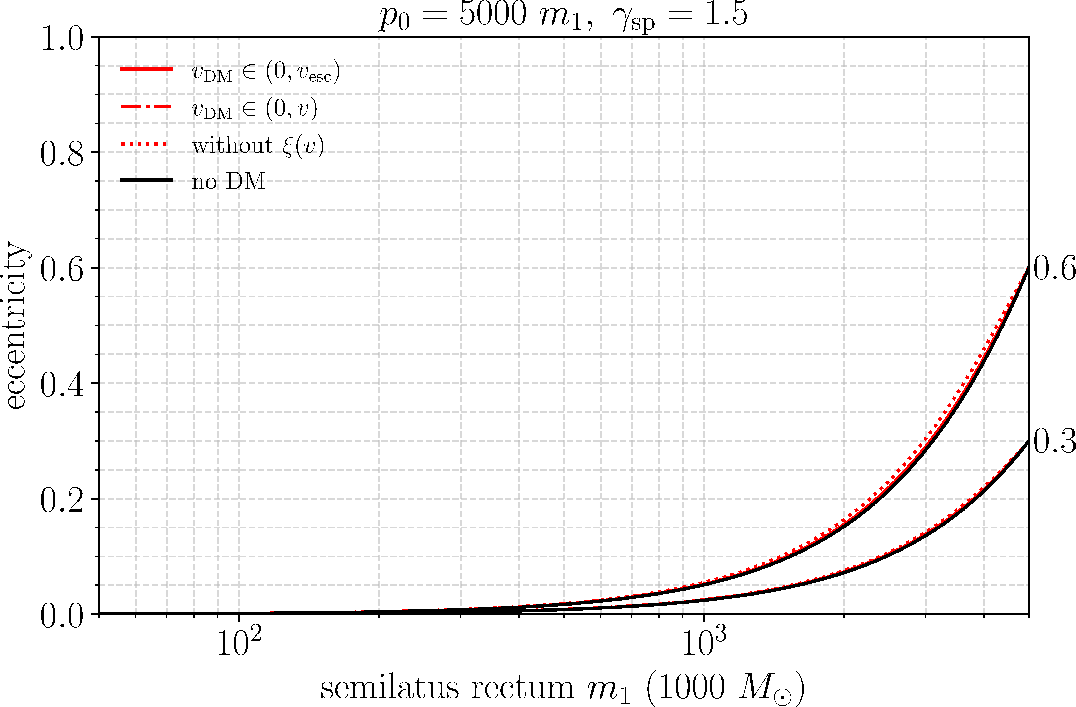}
        \includegraphics[width=0.3\textwidth]{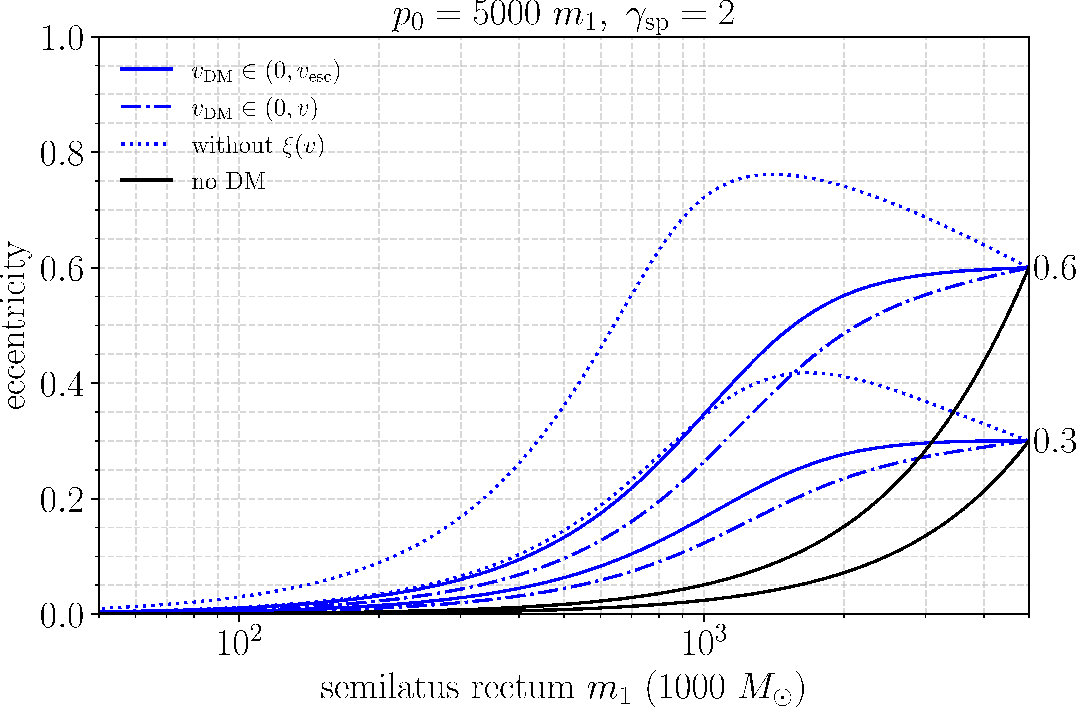}
        \includegraphics[width=0.3\textwidth]{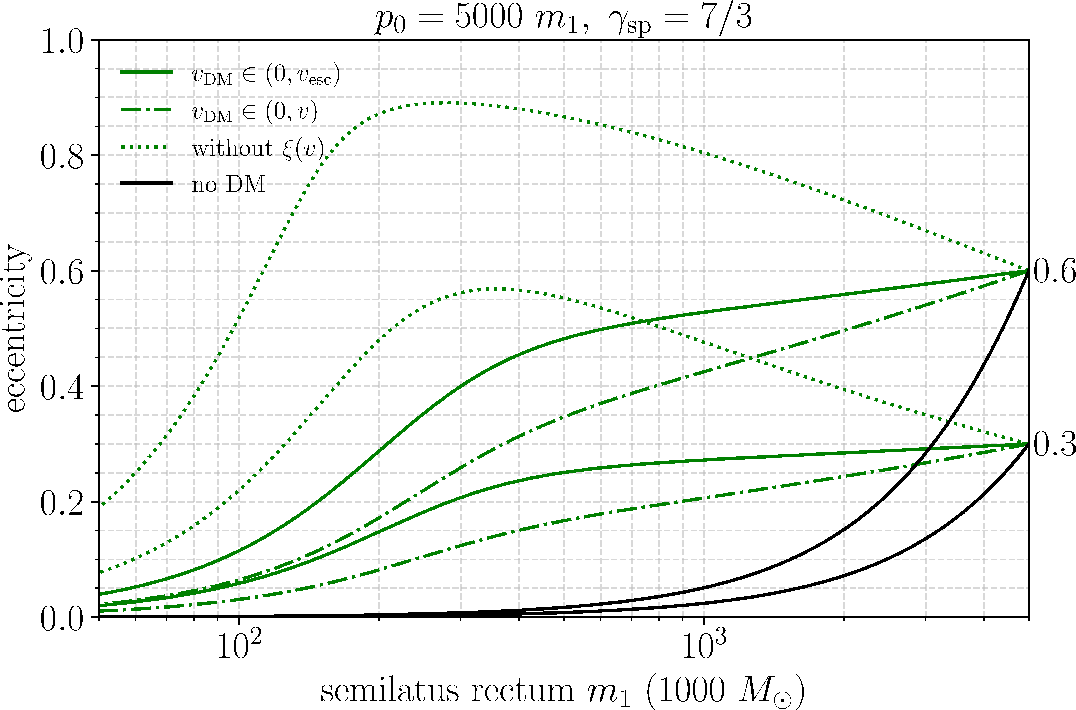}\\
        \includegraphics[width=0.3\textwidth]{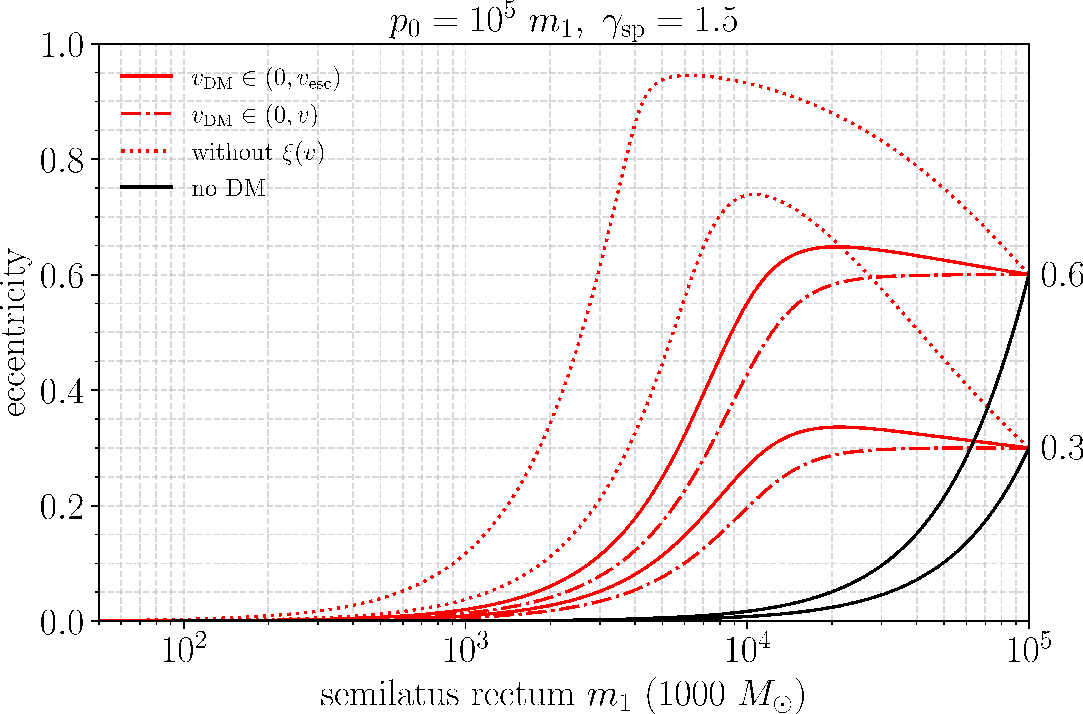}
        \includegraphics[width=0.3\textwidth]{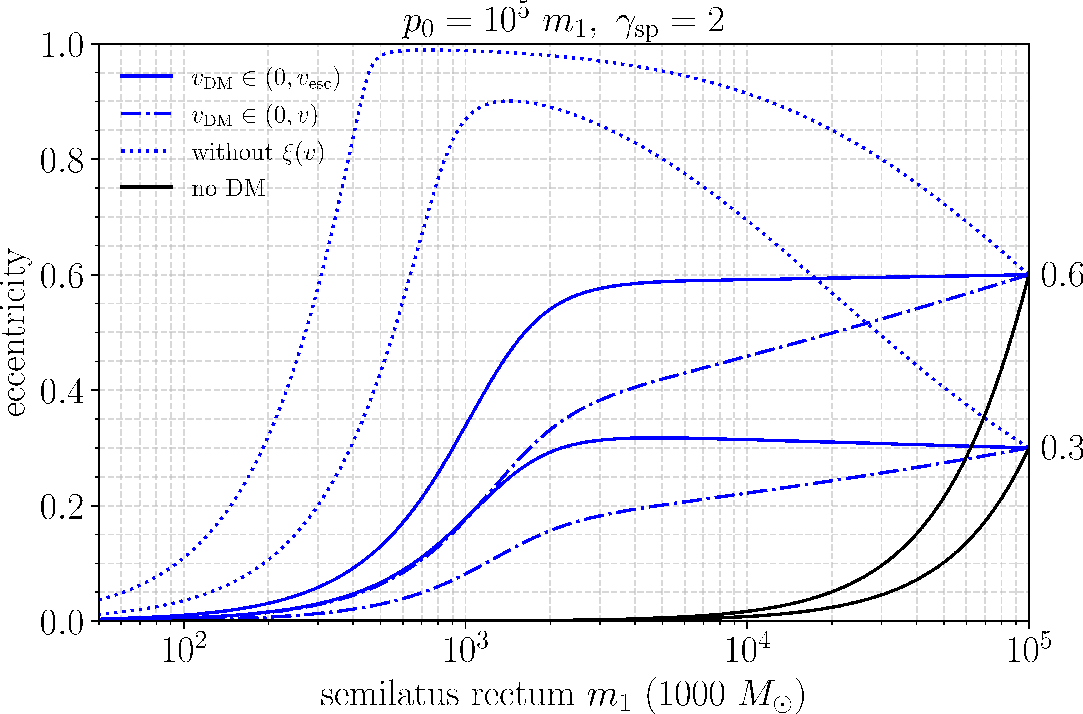}
        \includegraphics[width=0.3\textwidth]{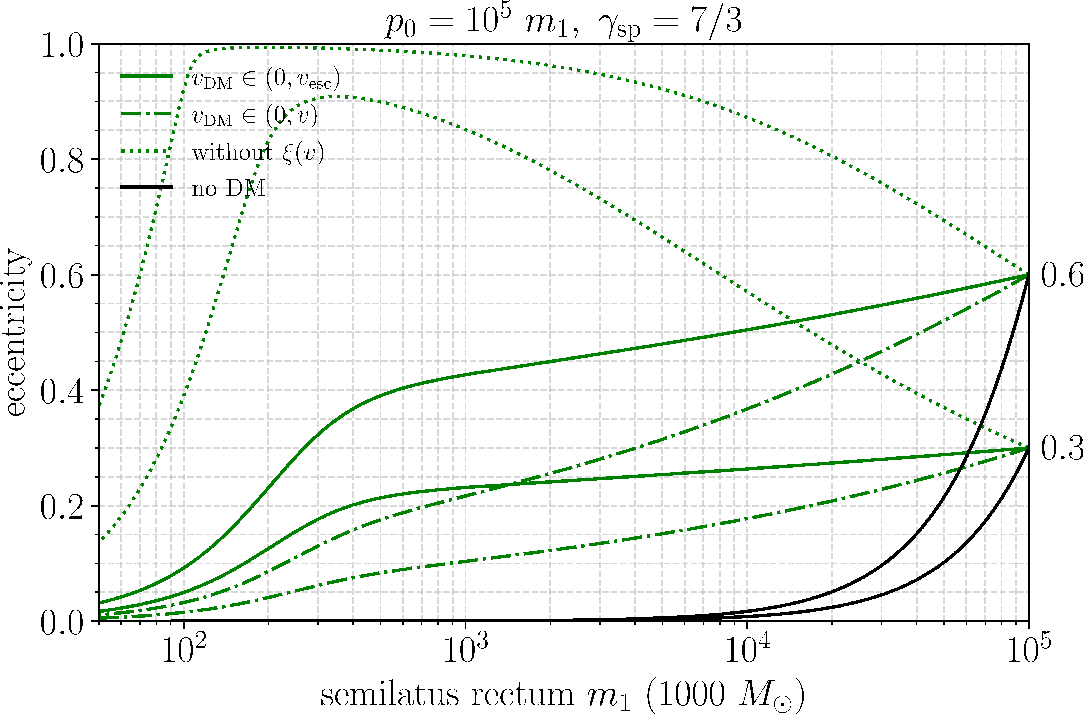}   
    \caption{The evolution of eccentricity $e$ as a function of semilatus rectum $p$. The dot lines correspond to cases without phase space description. The dashdot lines correspond to the cases without the contribution of faster moving DM particles. The solid lines correspond to cases including the contribution of the faster moving DM particles. The black lines correspond to cases without DM. The red, blue and green lines correspond to a spike power law index $\gamma_\mathrm{sp}=1.5$,$2$ and $7/3$, respectively. Top panels: The evolution of $e$ with initial $p=5000\ m_1$. Lower panels: The evolution of $e$ with initial $p=10^5\ m_1$.}
    \label{fig:ep}
\end{figure*}

The evolution of the eccentricity $e$ as a function of semilatus rectum $p=a(1-e^2)$ is plotted in Fig. \ref{fig:ep}. It is found that in the cases without phase space description, the dynamical friction dominates tends to eccentricify the orbits. Thus in the earlier stage of orbital evolution, the eccentricity increases due to the effects of dynamical friction. And in the later stage of orbital evolution, the GW emission dominates the orbital evolution, and the eccentricity decreases, as is described in \cite{PhysRevD.100.043013}.

In the cases without the contribution of the faster moving DM particles, the dynamical friction tends to circularize the orbit, thus the eccentricity decreases monotonically. In the later stage of orbital evolution, the energy loss is dominated by the GW emission and the eccentricity decreases faster, as is described in \cite{PhysRevD.105.063029}.

In the cases including the contribution of the faster moving DM particles, the evolution of eccentricity depends on the power law index of DM minispike $\gamma_\mathrm{sp}$. We numerically find the transition from increasing the eccentricity to decreasing the eccentricity occurs at $\gamma_\mathrm{sp}\approx2$. For $\gamma_\mathrm{sp}\lesssim2$, the dynamical friction tends to eccentricify the orbits. Thus the eccentricity increases in the earlier stage of orbital evolution. 

The difference in these three cases is due to the different behaviours of $\xi(v)$. \cite{PhysRevD.105.063029} proves that the dynamical friction due to the slow moving DM particles will circularize the orbit, $\left(\frac{de}{dt}\right)_\mathrm{slow}<0$. Our calculation proves that the faster moving DM particles will eccentricify the orbit, $\left(\frac{de}{dt}\right)_\mathrm{fast}>0$ (see Appendix \ref{sec:A1}). Therefore, the evolution of the eccentricity depends on the competition between the fast moving DM particles and the slow moving DM particles.

We next investigate the time evolution of semimajor axis $a$ and eccentricity $e$ under the dynamical friction in the three cases. The initial semilatus rectum $p$ is set to $p_0 = 5000, 10^5 \ m_1$ respectively and the rest of the parameters remain the same.

\begin{figure*}
        \includegraphics[width=0.45\textwidth]{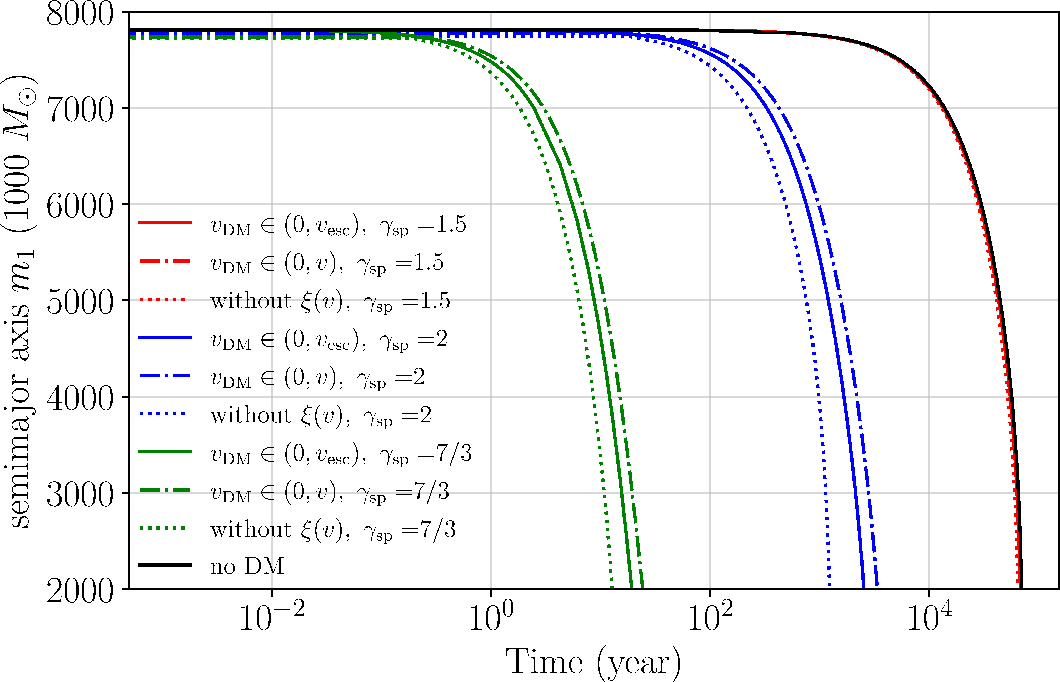}
        \includegraphics[width=0.45\textwidth]{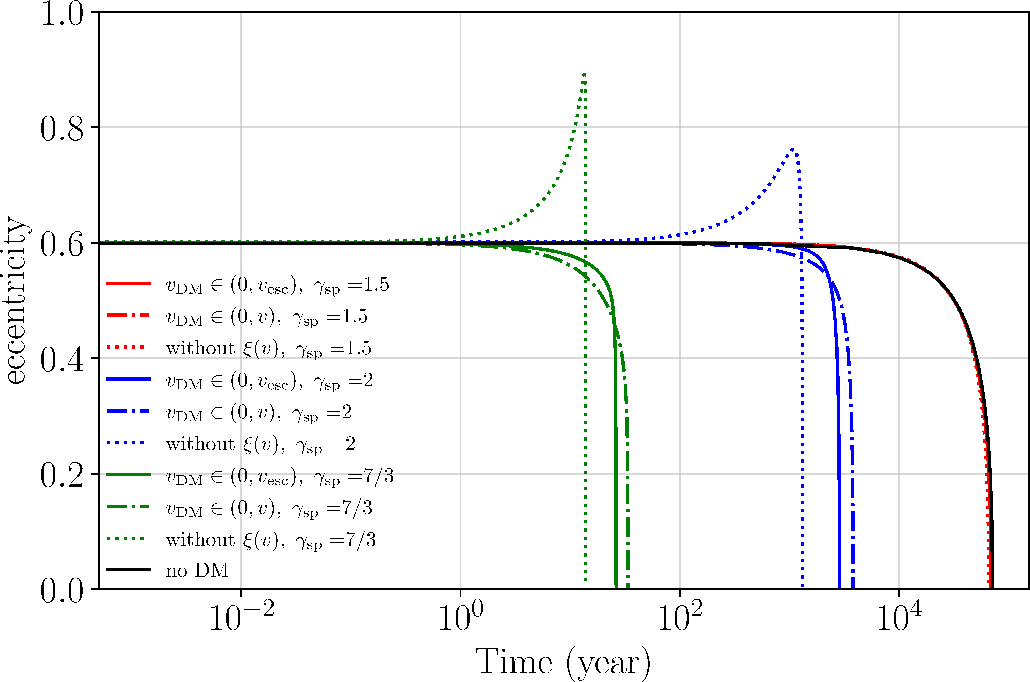}\\
        \includegraphics[width=0.45\textwidth]{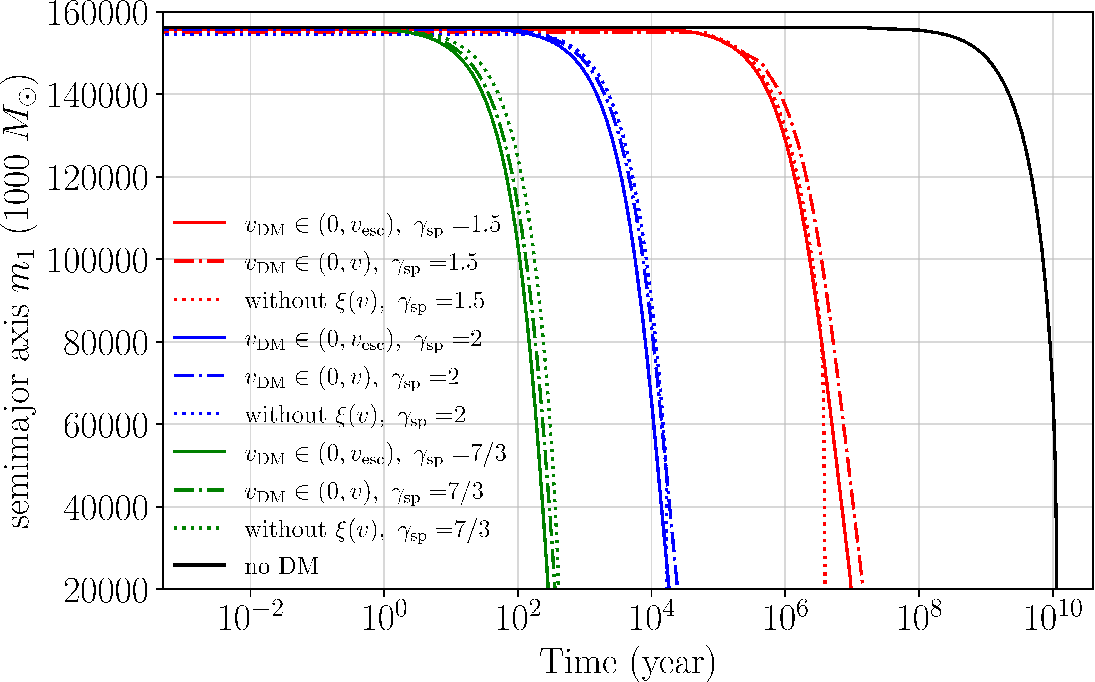}
        \includegraphics[width=0.45\textwidth]{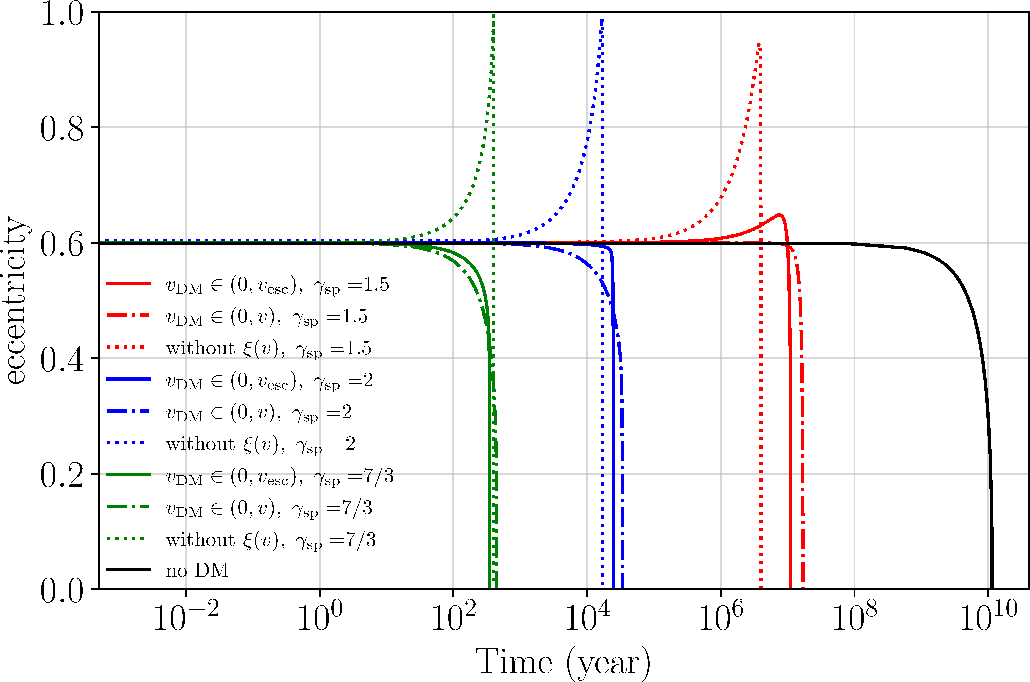}
    \caption{The evolution of semimajor axis $a$ and eccentricity $e$ as functions of time $t$. The dot lines correspond to cases without phase space description. The dashdot lines correspond to cases without faster moving DM particles. The solid lines correspond to cases including the contribution of the faster moving DM particles. The black lines correspond to cases without DM halo. The red, blue and green lines correspond to a spike power law index $\gamma_\mathrm{sp}=1.5$,$2$ and $7/3$, respectively. Top panels: The evolution of $a$ and $e$ with initial $p=5000\ m_1$. Lower panels: The evolution of $a$ and $e$ with initial $p=10^5\ m_1$.}
    \label{fig:atet}
\end{figure*}

The left panels of Fig. \ref{fig:atet} show that the existence of DM minispikes will make semimajor axis $a$ decrease faster and the inspiral time therefore shortens significantly. The right panels show that in the cases without $\xi(v)$, the eccentricity increases in the earlier inspiral stage when the energy loss is dominated by the dynamical friction. In the cases without the faster moving DM particles, the eccentricity decreases monotonically. In the cases with faster moving DM particles, The eccentricity increases in the earlier inspiral stage with $\gamma_\mathrm{sp}=1.5$, which makes a significantly difference with the previous cases.

\subsection{Dynamical friction effects on GW characteristic strain}
\label{s3B}
In this section we focus on the dynamical friction effects on the gravitational characteristic strain.

\begin{figure*}
        \includegraphics[width=0.32\textwidth]{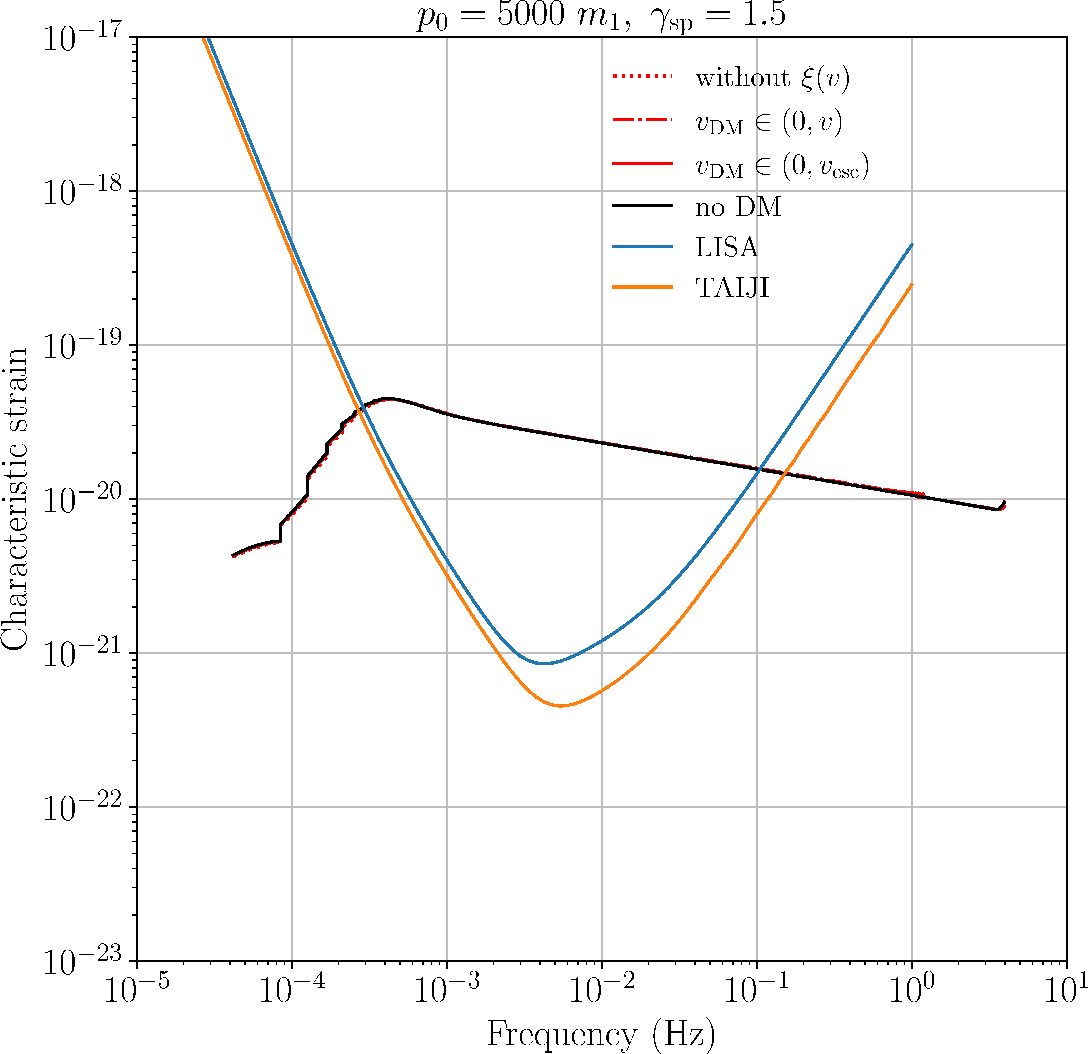}
        \includegraphics[width=0.32\textwidth]{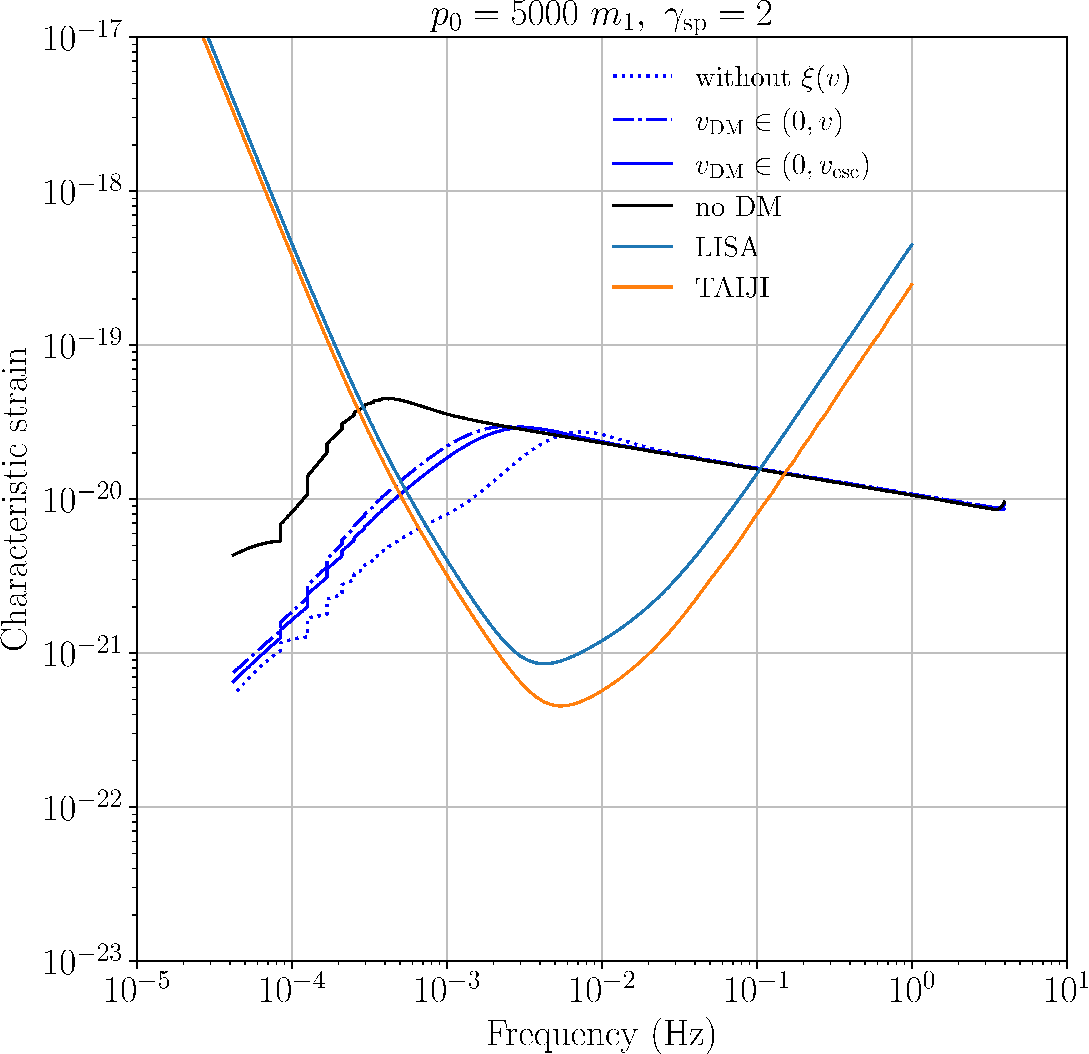}
        \includegraphics[width=0.32\textwidth]{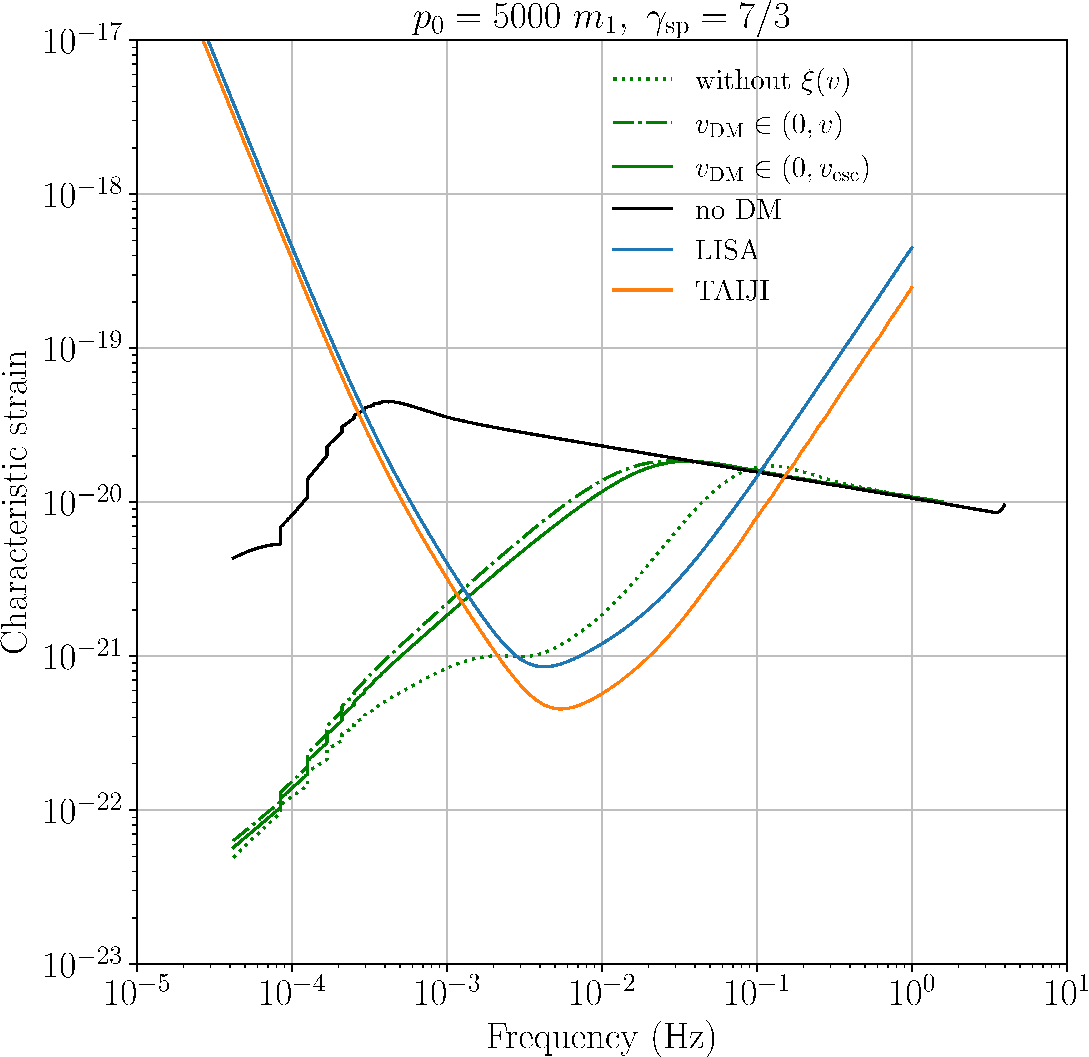}
    \caption{The effects of dynamical friction on the characteristic strain $h^{(2)}_{+}$. The dot lines correspond to cases without phase space description. The dashdot lines correspond to cases without faster moving DM particles. The solid lines correspond to cases including the contribution of the faster moving DM particles. The black lines correspond to cases without DM. The red, blue and green lines correspond to a spike power law index $\gamma_\mathrm{sp}=1.5$,$2$ and $7/3$, respectively.}
    \label{fig:gw_strain}
\end{figure*}

The characteristic strains of GW signals and the sensitivity curves of LISA and TAIJI are plotted in Fig. \ref{fig:gw_strain}.
The initial semilatus rectum $p$ is set to $p_0 = 5000 \ m_1$, the eccentricity $e$ is set to $e_0 = 0.6$, the luminosity distance $d_L$ is set to $d_L=500\ \text{Mpc}$ and the rest of the parameters remain the same. 

It is shown that the GW characteristic strains in three cases are suppressed at lower frequency domain where the evolution is dominated by the dynamical friction, as is described in \cite{Li:2021pxf}. In the higher frequency domain, the evolution is dominated by the GW emission, therefore there is no significant differences between the cases with and without the dynamical friction. The enhancement of characteristic in the higher domain found in \cite{Li:2021pxf} is also insignificant in our cases. 

\begin{figure}
    \centering
    \includegraphics[width=0.5\textwidth]{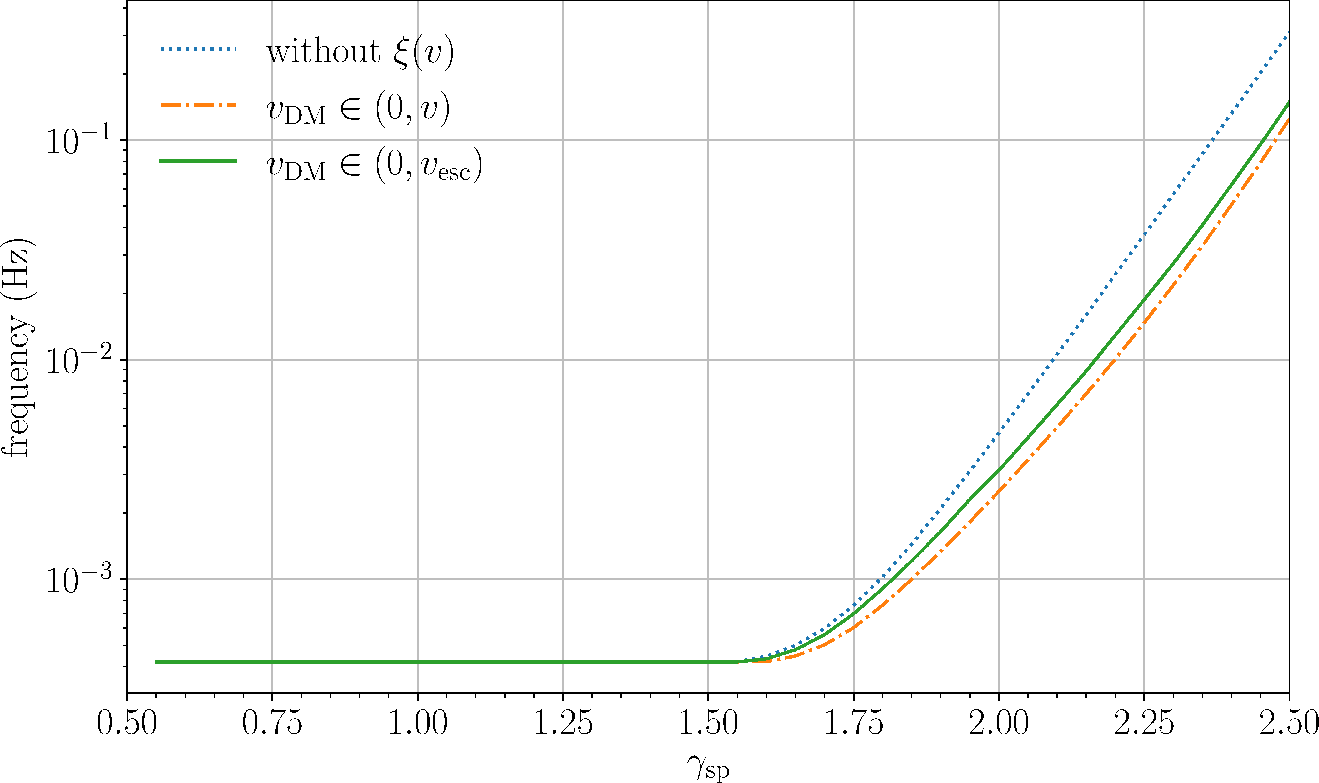}
    \caption{The relation between the frequency peak values of GW characteristic strains and power law index of DM spike $\gamma_\mathrm{sp}$.The dot lines correspond to cases without phase space description. The dashdot lines correspond to the cases without the contribution of faster moving DM particles. The solid lines correspond to cases including the contribution of the faster moving DM particles.}
    \label{fig:fgamma}
\end{figure}

Furthermore, It is found that the suppression effect is weak when $\gamma_\mathrm{sp}$ is small, and it become stronger as the power law index of DM spike $\gamma_\mathrm{sp}$ increases. 
Correspondingly, the maximum characteristic strains $h_c^{\mathrm{max}}$ also occur at higher frequencies as $\gamma_\mathrm{sp}$ increases. The relation between frequencies of maximum characteristic strain $f\left(h_c^{\mathrm{max}}\right)$  and $\gamma_\mathrm{sp}$ is shown in Fig. \ref{fig:fgamma}.

It is shown that when $\gamma_\mathrm{sp}$ is small, the effects of dynamical friction are insignificant, thus $f\left(h_c^{\mathrm{max}}\right)$ do not change significantly, and the differences between the case without DM and the cases with DM is not obvious. However, when $\gamma_\mathrm{sp}$ is large, the effects of dynamical friction are significant, thus $f\left(h_c^{\mathrm{max}}\right)$ enhance significantly, and it is possible to distinguish the case without DM and the cases with DM.
In order to distinguish differences between the cases with and without DM, we derive a general relation between the $f\left(h_c^{\mathrm{max}}\right)$ and $\gamma_\mathrm{sp}$.

\begin{equation}
\begin{aligned}
    \label{eq15}
    \frac{f\left(h_c^{\mathrm{max}}\right)}{f_0}&=\left[X(0)Y(0)\right]^{\frac{3}{11}}\left[X(e)Y(\beta(\gamma_\mathrm{sp}))\right]^{-\frac{3}{11-2\gamma_\mathrm{sp}-\beta}}\\
    &\sim Y(0)^{\frac{3}{11}}Y(\beta(\gamma_\mathrm{sp}))^{-\frac{3}{11-2\gamma_\mathrm{sp}-\beta}}
\end{aligned}
\end{equation}
Where $f_0$ is the frequency of the maximum character strain without DM, $\beta$ is the power index of phase space factor $\xi(v)$ (see Appendix \ref{sec:A1}), $X(e)$ and $Y(\beta(\gamma_\mathrm{sp}))$ are defined in Appendix \ref{sec:A2}.

The enhancement of $f\left(h_c^{\mathrm{max}}\right)$ can be explained through this equation. For example, we choose power law indices $\gamma_\mathrm{sp}=1.5,\ 2$ and the corresponding $\beta(\gamma_\mathrm{sp})$ are $3,\ 2.75$. The ratio of orbital frequencies are
\begin{align}
\label{eqB7}
&\frac{f_{\gamma_\mathrm{1.5}}}{f_0}\sim Y(0)^{\frac{3}{11}}Y(3)^{-\frac{3}{11-2\cdot1.5-3}}\approx0.06\\
\label{eqB8}
&\frac{f_{\gamma_\mathrm{2}}}{f_0}\sim Y(0)^{\frac{3}{11}}Y(2.75)^{-\frac{3}{11-2\cdot2-2.75}}\approx4.30
\end{align}
which corresponds well with the result in Fig. \ref{fig:gw_strain} and Fig. \ref{fig:fgamma}.

\section{conclusion}
\label{s5}
In this paper we consider the IMRIs surrounded by DM minispikes, and investigate the effects of general dynamical friction on the orbital evolution of IMRIs. Numerical calculations reveal that the evolution of eccentricity  is contingent upon the power law index of DM minispike $\gamma_\mathrm{sp}$. A transition point is identified at $\gamma_\mathrm{sp}\approx2$, beyond which the eccentricity evolution reverses from increase to decrease. 

For $\gamma_\mathrm{sp}\lesssim2$, the dynamical friction tends to increase the eccentricity. Furthermore, it is proved that the dynamical friction due to the faster moving DM particles will eccentricify the orbit, independently of $\gamma_\mathrm{sp}$, which is contrast to the result of dynamical friction due to the slower moving DM particles. Thus, the circularization and eccentricification of orbit depends on the fraction of slower and faster part of dynamical friction.

We also investigate the effects of general dynamical friction on GW characteristic strains. We find that GW characteristic strains are suppressed at lower frequency domain, where the evolution is dominated by the dynamical friction. It is also found that the peak values of characteristic strains occur at higher frequencies as the power law index of DM spike $\gamma_\mathrm{sp}$ increases. We derive a general relation between the frequency peak value of characteristic strain of GWs and the power law index of DM minispike $\gamma_\mathrm{sp}$ to explain this phenomenon. This relation can also help to confine the range of $\gamma_\mathrm{sp}$ in the future GW detection.
In summary, the general dynamical friction including the contribution of faster moving particles has significant effects on IMRIs, and it is expected that the future GW detectors will observes these effects. Using that analytical relations, the the presence of dark matter, as well as the dark matter halo density distribution, can be determined from the detected GW data in the future.

\section*{Acknowledgements}
Yu-Chen Zhou thanks Yuxin Liu for the helpful discussions.  
This work has been supported by the Fundamental Research Funds for the Central Universities.
This work has been supported in part by the Strategic Priority Research Program of the Chinese Academy of Sciences under Grant No. XDA15020701 and XDA15020708. In this paper, The numerical computation of this work was completed at TAIJI Cluster of University of Chinese Academy of Sciences. 

\appendix
\section{Dynamical friction effects on eccentricity evolution}\label{sec:A1}
The dynamical friction effects on eccentricity evolution can be seen from the time derivative of eccentricity $\frac{de}{dt}$ [Eq. (\ref{eq9})]. Disregarding the factors and the contribution from the GW emission, we obtain the eccentricity evolution dominated by dynamical friction
\begin{equation}
    \label{eqA1}
    \left(\frac{de}{dt}\right)_\mathrm{DF}\sim-\left(\frac{1}{E}\left<\frac{dE}{dt}\right>_\mathrm{DF}+\frac{2}{L}\left<\frac{dL}{dt}\right>_\mathrm{DF}\right)
\end{equation}

The energy and angular momentum loss due to dynamical friction are
\begin{equation}
\label{eqA2}
\begin{aligned}
\left<\frac{dE}{dt}\right>_\mathrm{DF}&=-\int^T_0{\frac{dt}{T}F_\mathrm{DF}v}\\
&=-(1-e^2)^\frac{3}{2}\int^{2\pi}_0{\frac{d\phi}{2\pi}\frac{F_\mathrm{DF}v}{(1+e\cos{\phi})^2}} 
\end{aligned}
\end{equation}
\begin{equation}
\label{eqA3}
\begin{aligned}
\left<\frac{dL}{dt}\right>_\mathrm{DF}&=-\sqrt{Ma(1-e^2)}\int^T_0{\frac{dt}{T}\frac{F_\mathrm{DF}}{v}}\\
&=-\sqrt{Ma}(1-e^2)^2\int^{2\pi}_0{\frac{d\phi}{2\pi}\frac{F_\mathrm{DF}}{(1+e\cos{\phi})^2 v}}
\end{aligned}
\end{equation}

With the expression of $E$, $L$, $\left<\frac{dE}{dt}\right>_\mathrm{DF}$ and $\left<\frac{dL}{dt}\right>_\mathrm{DF}$, The eccentricity evolution can be rewritten as \cite{PhysRevD.105.063029}
\begin{equation}
\begin{aligned}
\label{eqA4}
\left(\frac{de}{dt}\right)_\mathrm{DF} \sim-\frac{2(1-e^2)^\frac{3}{2}}{\mu}&\int_0^{2\pi} \frac{d\phi}{2\pi}\frac{F_\mathrm{DF}}{(1+e\cos\phi)^2}\\
&\ \times \left(\frac{av}{m}-\frac{1}{v}\right)
\end{aligned}
\end{equation}

The dynamical friction $F_\mathrm{DF}$ can be written in the form
\begin{equation}
\label{eqA5}
\begin{aligned}
F_\mathrm{DF}&=F_0r^{-\gamma_\mathrm{sp}}v^{-2}\xi(v)\\
&=\left\{
\begin{array}{ll}
F_0r^{-\gamma_\mathrm{sp}}b_\mathrm{slow}v^{\beta_\mathrm{slow}-2}, &v_\mathrm{DM}<v\\
F_0r^{-\gamma_\mathrm{sp}}b_\mathrm{fast}(v_\mathrm{esc}-v)^{\beta_\mathrm{fast}}v^{-2}, &v_\mathrm{DM}>v
\end{array}
\right.
\end{aligned}
\end{equation}
where we adopt the assumption that the phase space factor $\xi(v)$ has a power law form \cite{PhysRevD.105.063029}. The two factors $b_\mathrm{slow}$ and $b_\mathrm{fast}$, and two power law indices $\beta_\mathrm{slow}$ and $\beta_\mathrm{fast}$ can be determined numerically through Eqs. (\ref{eq2}) and (\ref{eq5}). The factor $F_0$ is
\begin{equation}
\label{eqA6}
F_0=4\pi m_2^2\rho_\mathrm{sp}r_\mathrm{sp}^{\gamma_\mathrm{sp}}
\end{equation}

The radius of the binary system $r$ and the velocity of the secondary BH $v$ can be expressed as functions of true anomaly $\phi$
\begin{equation}
\label{eqA7}
r=\frac{a(1-e^2)}{1+e\cos{\phi}}
\end{equation}
\begin{equation}
\label{eqA8}
v=\sqrt{\frac{2M}{r}-\frac{M}{a}}=\sqrt{\frac{M(e^2+2e\cos{\phi}+1)}{a(1-e^2)}}
\end{equation}

Substituting Eqs. (\ref{eqA5}),(\ref{eqA7}) and (\ref{eqA8}), we can rewrite the eccentricity evolution as the sum of two parts
\begin{equation}
\label{eqA9}
\left(\frac{de}{dt}\right)_\mathrm{DF}=\left(\frac{de}{dt}\right)_{\mathrm{DF,slow}}+\left(\frac{de}{dt}\right)_{\mathrm{DF,fast}}
\end{equation}

Disregarding the factors, the expression of $\left(\frac{de}{dt}\right)_{\mathrm{DF,slow}}$ and $\left(\frac{de}{dt}\right)_{\mathrm{DF,fast}}$ are
\begin{equation}
\label{eqA10}
\begin{aligned}
\left(\frac{de}{dt}\right)_{\mathrm{slow}} &\sim-\int_0^{2\pi} \frac{d\phi}{2\pi}(e+\cos\phi)(1+e \cos\phi)^{\gamma_\mathrm{sp}-2}\\
&\qquad \times v^{\beta_\mathrm{slow}-3}\\
&=-\int_0^{2\pi} \frac{d\phi}{2\pi} (e+\cos\phi)(1+e \cos\phi)^{\gamma_\mathrm{sp}-2}\\
&\qquad \times\left(e^2+2e\cos{\phi}+1\right)^{(\beta_{\mathrm{slow}}-3)/2}
\end{aligned}
\end{equation}

\begin{equation}
\label{eqA11}
\begin{aligned}
\left(\frac{de}{dt}\right)_{\mathrm{fast}} &\sim-\int_0^{2\pi} \frac{d\phi}{2\pi}(e+\cos\phi)(1+e \cos\phi)^{\gamma_\mathrm{sp}-2}\\
&\qquad \times \left(v_\mathrm{esc}-v\right)^{\beta_\mathrm{fast}}v^{-3}
\end{aligned}
\end{equation}

To first order in $e$, the integral of the slow part can be expressed as
\begin{equation}
\label{eqA12}
\begin{aligned}
\left(\frac{de}{dt}\right)_{\mathrm{slow}} &\sim-\int_0^{2\pi} \frac{d\phi}{2\pi} \left[\cos{\phi}+e+(\beta_{\mathrm{slow}}+\gamma_\mathrm{sp}-5)e\cos^2{\phi}\right]\\
&=-\frac{e}{2}\left(\gamma_\mathrm{sp}+\beta_\mathrm{slow}-3\right)
\end{aligned}
\end{equation}
The sign of the expression depends on the $\gamma_\mathrm{sp}$ and $\beta_\mathrm{slow}$. Numerically we find that when $\gamma_\mathrm{sp}$ increases from 0.5 to 3, $\beta_\mathrm{slow}$ decreases from 3.55 to 2.35. Therefore, $\left(de/dt\right)_{\mathrm{slow}}<0$, the dynamical friction caused by DM particles moving slower than the secondary BH tends to decrease the eccentricity of the orbit, which agrees with the conclusion in \cite{PhysRevD.105.063029}.

We next compute the integral of the fast part in the same approach
\begin{equation}
\label{eqA13}
\begin{aligned}
\left(\frac{de}{dt}\right)_{\mathrm{fast}} &\sim-\int_0^{2\pi} \frac{d\phi}{2\pi} \bigg[(\sqrt{2}-1)^{\beta_\mathrm{fast}}\cos{\phi}+\left(\sqrt{2}-1\right)^{\beta_\mathrm{fast}}e\\
&\quad+(\sqrt{2}-1)^{\beta_\mathrm{fast}}\left(\gamma_\mathrm{sp}-\frac{\sqrt{2}}{2}\beta-5\right)e\cos^2{\phi}\bigg]\\
&\sim-\int_0^{2\pi} \frac{d\phi}{2\pi}\left[\cos{\phi}+e+\left(\gamma_\mathrm{sp}-\frac{\sqrt{2}}{2}\beta-5\right)e\cos^2{\phi}\right]\\
&=-\frac{e}{2}\left(\gamma_\mathrm{sp}-\frac{\sqrt{2}}{2}\beta_\mathrm{fast}-3\right)
\end{aligned}
\end{equation}
The sign of the expression depends on the $\gamma_\mathrm{sp}$ and $\beta_\mathrm{fast}$. Numerically we find that when $\gamma_\mathrm{sp}$ increases from 0.5 to 3, $\beta_\mathrm{slow}$ decreases from 1.37 to 0.42. Therefore, $\left(de/dt\right)_{\mathrm{fast}}>0$, the dynamical friction caused by DM particles moving faster than the secondary BH tends to increase the eccentricity of the orbit.

\section{THE RELATIONSHIP BETWEEN POWER LAW INDEX OF DM HALO AND THE MAXIMUM VALUE OF CHARACTERISTIC STRAIN}
\label{sec:A2}
The peak value of characteristic strain occurs at the transition stage where the GW emission dominates the orbital evolution. 

To simplify the calculation, we assume the transition stage mentioned in Sec. \ref{s3B} is that the energy loss of dynamical friction and GW emission over one obrit is equal
\begin{equation}
\label{eqB1}
\left<\Delta E\right>_\mathrm{DF}=\left<\Delta E\right>_\mathrm{GW}
\end{equation}
Furthermore, we only consider the slow moving DM particles, since the difference between this case and the case with fast moving DM particles in Fig. \ref{fig:gw_strain} and Fig. \ref{fig:fgamma} is not obvious. Substituting the energy loss of dynamical friction and GW emission, we obtain
\begin{equation}
\label{eqB2}
F_0\left(\frac{r_\mathrm{sp}}{r}\right)^{\gamma_\mathrm{sp}}v^{\beta-1}=\frac{32}{5}\frac{\mu^2 M^3}{r^5}X(e)
\end{equation}
where the function X(e) is
\begin{equation}
\label{eqB3}
X(e)=\frac{1+\frac{73}{24}e^2+\frac{37}{96}e^4}{(1+e \cos{\phi})^5}(1-e^2)^{\frac{3}{2}}
\end{equation}
The radius $r$ can be solved from this equation
\begin{equation}
\label{eqB4}
r=r_\mathrm{sp}\left(X(e)Y(\beta(\gamma_\mathrm{sp}))\right)^{\frac{1}{\frac{11}{2}-\gamma_\mathrm{sp}-\frac{\beta}{2}}}
\end{equation}
where the function $Y(\beta(\gamma_\mathrm{sp}))$ is
\begin{equation}
\label{eqB5}
Y(\beta(\gamma_\mathrm{sp}))=\frac{32}{5}\frac{\mu^2 M^3}{r_\mathrm{sp}^5}\left(\frac{r_\mathrm{sp}}{4\pi m_2^2\rho_\mathrm{sp}M}\right)^{\frac{\beta-1}{2}}
\end{equation}
The frequency is proportional to $r^{-\frac{3}{2}}$, thus the ratio of orbital frequencies with different power law indices $\gamma_\mathrm{sp}$ is
\begin{equation}
\label{eqB6}
\frac{f}{f_0}=\left(\frac{r}{r0}\right)^{-\frac{3}{2}}\sim Y(0)^{\frac{3}{11}}Y(\beta(\gamma_\mathrm{sp}))^{-\frac{3}{11-2\gamma_\mathrm{sp}-\beta}}
\end{equation}
where $f_0$ is the peak value of characteristic strain in the cases without DM, and $r_0$ is the corresponding radius in these cases.
 The $X(e)$ term can be omitted when eccentricity is small.
\bibliographystyle{aasjournalv7}
\bibliography{citations,library}
\end{document}